
\input harvmac

\def\frac#1#2{{#1\over#2}}

\def\journal#1&#2(#3){\unskip, #1~\bf #2 \rm(19#3) }
\def\andjournal#1&#2(#3){\sl #1~\bf #2 \rm (19#3) }

\catcode`\@=11\def\slash#1{\mathord{\mathpalette\c@ncel{#1}}}
\overfullrule=0pt
\def\steepslash{\c@ncel}
\def\frac#1#2{{#1\over #2}}

\def\:{\!:\!}
\def\inbar{\,\vrule height1.5ex width.4pt depth0pt}
\def\IQ{\relax\,\hbox{$\inbar\kern-.3em{\rm Q}$}}
\def\IB{\relax{\rm I\kern-.18em B}}
\def\IC{\relax\hbox{$\inbar\kern-.3em{\rm C}$}}
\def\IP{\relax{\rm I\kern-.18em P}}
\def\IR{\relax{\rm I\kern-.18em R}}
\def\ZZ{\relax\ifmmode\mathchoice
{\hbox{Z\kern-.4em Z}}{\hbox{Z\kern-.4em Z}}
{\lower.9pt\hbox{Z\kern-.4em Z}}
{\lower1.2pt\hbox{Z\kern-.4em Z}}\else{Z\kern-.4em Z}\fi}

\catcode`\@=12

\def\npb#1(#2)#3{{ Nucl. Phys. }{B#1} (#2) #3}
\def\plb#1(#2)#3{{ Phys. Lett. }{#1B} (#2) #3}
\def\pla#1(#2)#3{{ Phys. Lett. }{#1A} (#2) #3}
\def\prl#1(#2)#3{{ Phys. Rev. Lett. }{#1} (#2) #3}
\def\mpla#1(#2)#3{{ Mod. Phys. Lett. }{A#1} (#2) #3}
\def\ijmpa#1(#2)#3{{ Int. J. Mod. Phys. }{A#1} (#2) #3}
\def\cmp#1(#2)#3{{ Comm. Math. Phys. }{#1} (#2) #3}
\def\cqg#1(#2)#3{{ Class. Quantum Grav. }{#1} (#2) #3}
\def\jmp#1(#2)#3{{ J. Math. Phys. }{#1} (#2) #3}
\def\anp#1(#2)#3{{ Ann. Phys. }{#1} (#2) #3}
\def\prd#1(#2)#3{{ Phys. Rev. } {D{#1}} (#2) #3}
\def\ptp#1(#2)#3{{ Progr. Theor. Phys. }{#1} (#2) #3}
\def\aom#1(#2)#3{{ Ann. Math. }{#1} (#2) #3}

\def\R{{\bf R}}
\def\Z{{\bf Z}}
\def\P{{\bf P}}
\def\Q{{\bf Q}}
\def\Z{{\bf Z}}
\def\cO{{\cal O}}

\def\cC{{\cal C}}

\def\cK{{\cal K}}

\def\cicy#1(#2|#3)#4{\left(\matrix{#2}\right|\!\!
                     \left|\matrix{#3}\right)^{{#4}}_{#1}}

\def\ra{\rightarrow}

\def\bs{\bigskip}
\def\qed{{\bf $\bullet$}}

\global\newcount\thmno \global\thmno=0
\def\question#1{\global\advance\thmno by1
\bigskip\noindent{\bf Question \secsym\the\thmno. }{\it #1}
\par\nobreak\medskip\nobreak}
\def\theorem#1{\global\advance\thmno by1
\bigskip\noindent{\bf Theorem \secsym\the\thmno. }{\it #1}
\par\nobreak\medskip\nobreak}
\def\proposition#1{\global\advance\thmno by1
\bigskip\noindent{\bf Proposition \secsym\the\thmno. }{\it #1}
\par\nobreak\medskip\nobreak}
\def\corollary#1{\global\advance\thmno by1
\bigskip\noindent{\bf Corollary \secsym\the\thmno. }{\it #1}
\par\nobreak\medskip\nobreak}
\def\lemma#1{\global\advance\thmno by1
\bigskip\noindent{\bf Lemma \secsym\the\thmno. }{\it #1}
\par\nobreak\medskip\nobreak}
\def\conjecture#1{\global\advance\thmno by1
\bigskip\noindent{\bf Conjecture \secsym\the\thmno. }{\it #1}
\par\nobreak\medskip\nobreak}
\def\exercise#1{\global\advance\thmno by1
\bigskip\noindent{\bf Exercise \secsym\the\thmno. }{\it #1}
\par\nobreak\medskip\nobreak}
\def\remark#1{\global\advance\thmno by1
\bigskip\noindent{\bf Remark \secsym\the\thmno. }{\it #1}
\par\nobreak\medskip\nobreak}
\def\problem#1{\global\advance\thmno by1
\bigskip\noindent{\bf Problem \secsym\the\thmno. }{\it #1}
\par\nobreak\medskip\nobreak}
\def\proof{\noindent Proof: }

\def\thmlab#1{\xdef #1{\the\thmno}\writedef{#1\leftbracket#1}\wrlabeL{#1=#1}}
%
%
\def\eqnres@t{\global\thmno=0%
\xdef\secsym{\the\secno.}\global\meqno=1\bigbreak\bigskip}
\def\sequentialequations{\def\eqnres@t{\bigbreak}}\xdef\secsym{}
\def\fivepoint{\def\rm{\fam0\fiverm}
\textfont0=\fiverm \scriptfont0=\fiverm \scriptscriptfont0=\fiverm
\textfont1=\fivei \scriptfont1=\fivei \scriptscriptfont1=\fivei
\textfont2=\fivesy \scriptfont2=\fivesy \scriptscriptfont2=\fivesy
\textfont\itfam=\fivei \def\it{\fam\itfam\fiveit}\def\sl{\fam\slfam\fivesl}%
\textfont\bffam=\fivebf \def\bf{\fam\bffam\fivebf}\rm}
\def\npb#1(#2)#3{{ Nucl. Phys. }{B#1} (#2) #3}
\def\plb#1(#2)#3{{ Phys. Lett. }{#1B} (#2) #3}
\def\pla#1(#2)#3{{ Phys. Lett. }{#1A} (#2) #3}
\def\prl#1(#2)#3{{ Phys. Rev. Lett. }{#1} (#2) #3}
\def\mpla#1(#2)#3{{ Mod. Phys. Lett. }{A#1} (#2) #3}
\def\ijmpa#1(#2)#3{{ Int. J. Mod. Phys. }{A#1} (#2) #3}
\def\cmp#1(#2)#3{{ Commun. Math. Phys. }{#1} (#2) #3}
\def\cqg#1(#2)#3{{ Class. Quantum Grav. }{#1} (#2) #3}
\def\jmp#1(#2)#3{{ J. Math. Phys. }{#1} (#2) #3}
\def\anp#1(#2)#3{{ Ann. Phys. }{#1} (#2) #3}
\def\prd#1(#2)#3{{ Phys. Rev.} {D\bf{#1}} (#2) #3}

\def\R{{\bf R}}
\def\Z{{\bf Z}}
\def\P{{\bf P}}
\def\Q{{\bf Q}}

\def\inbar{\,\vrule height1.5ex width.4pt depth0pt}
\def\IQ{\relax\,\hbox{$\inbar\kern-.3em{\rm Q}$}}
\def\IB{\relax{\rm I\kern-.18em B}}
\def\IC{\relax\hbox{$\inbar\kern-.3em{\rm C}$}}
\def\IP{\relax{\rm I\kern-.18em P}}
\def\IR{\relax{\rm I\kern-.18em R}}
\def\ZZ{\relax\ifmmode\mathchoice
{\hbox{Z\kern-.4em Z}}{\hbox{Z\kern-.4em Z}}
{\lower.9pt\hbox{Z\kern-.4em Z}}
{\lower1.2pt\hbox{Z\kern-.4em Z}}\else{Z\kern-.4em Z}\fi}

\Title{}{Mirror Maps, Modular Relations and Hypergeometric Series I
\footnote{${}^\diamond$}{Research supported by grant DE-FG02-88-ER-25065.}}

\centerline{
Bong H. Lian$^{1,2}$\footnote{}{$^1$~~Department of Mathematics,
Brandeis University, Waltham, MA 02154.}
 and Shing-Tung Yau$^2$\footnote{}{$^2$~~Department of Mathematics,
Harvard University, Cambridge, MA 02138.} }

\vskip .2in

Abstract. Motivated by the recent work of Kachru-Vafa in string theory,
we study in Part A of this paper, certain identities involving modular forms,
 hypergeometric series, and more generally series solutions to Fuchsian
 equations. The identity which arises in string theory is
the simpliest of its kind. There are nontrivial
generalizations of the identity which appear new. We give many such examples
-- all of which arise in mirror symmetry for algebraic K3 surfaces.
 In Part B, we study the integrality property of
certain $q$-series, known as mirror maps, which arise in mirror symmetry.

\Date{hep-th/9507151} 


\lref\vk{C. Vafa and S. Kachru, {\sl Exact Results
for N=2 Compactifications of Heterotic Strings}, hep-th/9505105.}
\lref\ss{J. Schwarz and A. Sen, {\sl Duality Symmetries of 4d Heterotic
Strings}, Phys. Lett. B312 (1993) 105.}
\lref\hulltownsend{C. Hull and P. Townsend,
{\sl Unity of Superstring Dualities}, hep-th/9410167.}
\lref\seibergwitten{N. Seiberg and E. Witten,
{\sl Electric-Magnetic Duality, Monopole Condensation, and Confinement
in $N=2$ Supersymmetric Yang-Mills Theory}, Nucl. Phys. B426 (1994) 19.}
\lref\witten{E. Witten, {\sl String Theory Dynamics in Various Dimensions},
hep-th/9503124.}
\lref\fhsv{S. Ferrara, J. Harvey, A. Strominger and C. Vafa,
{\sl Second-Quantized Mirror Symmetry}, hep-th/9505162.}
\lref\vafawitten{C. Vafa and E. Witten, {\sl Dual String Pairs with
$N=1$ and $N=2$ Supersymmetry in Four Dimensions}, hep-th/9507050.}

\newsec{Introduction}

Physicists have recently come up with apparently different
realizations of a same physical theory, using
a remarkable phenomenon known as {\it duality} \ss\hulltownsend\seibergwitten
\witten\fhsv\vafawitten.
In a given theory,  this allows one
to study and compute the same physical quantity by drastically
different means.
 Upon proper mathematical
interpretation, such computations often lead to
remarkable relations between mathematical objects.
 In Part A of this paper,
we study certain functional relations, and their generalizations,
arising in the recent work of Kachru-Vafa on the so-called
heterotic-type II duality.
 The objects in question are certain
automorphic forms, and the relations involved are power series identities
which we loosely call ``modular relations''.

In Part B, we consider the mirror map for the configuration of degree $p$
Calabi-Yau hypersurfaces in $\P^{p-1}$ -- one for each odd prime $p$.
We will prove that the coefficients of the mirror map $z(q)$, as a q-series,
are integral. This problem goes back to the first celebrated
computation of the mirror map
for the quintic in $\P^4$ \ref\cdgp{P. Candelas, X. De la Ossa, P. Green and L.
Parkes, Nucl. Phys. B359 (1991) 21.}.
We have established for the first time the integrality of mirror maps for
infinitely many cases. While we restrict our detailed discussion to the
cases of the degree $p$ hypersurface in $\P^{p-1}$, we will later indicate how
the same technique can be applied to  other known cases as well.

The technique consists of some elementary applications of:
\item{1.} The exponential version of Dwork's lemma on p-adic power series;
\item{2.} Some theorems of Dwork on hypergeometric functions;
\item{3.} An estimate using the p-adic Gamma function.

 The relevance of Dwork's theory on the integrality question was suggested
by Ogus.\footnote{${}^1$}{We thank M. Kontsevich for communicating this idea to
us.}

{\it Notation:} Throughout this paper, a holomorphic function $f$ defined on a
disc $|q|<r$ will be written as $f(q)$ when regarded as a power series. We
denote $q{df\over dq}$ by $f'$. The parameter $q$ will sometimes be
identified with $e^{2\pi it}$ with $Im\ t>0$. We will say that $f$ is a
solution to a differential
operator $L$ if $Lf=0$ in a domain where this makes sense.

{\bf Acknowledgements.} We thank A.O.L. Atkin,
N. Elkies, B. Gross, A. Klemm, M. Kontsevich
and G. Zuckerman for helpful discussions.

\newsec{Part A. Modular Relations: the first example}

The identity in question is:
\eqn\one{\left(\sum_{n=0}^\infty{(6n)!\over(3n)!(n!)^3} {1\over
j(q)^n}\right)^2
= E_4(q).}
Here $j$ is the Dedekind-Klein j-function, $E_4$ is the Eisenstein series of
weight 4 with respect to $SL(2,\Z)$. It has a Fourier expansion
\eqn\dumb{E_4(q)=1+240\sum_{n=1}^\infty\sigma_3(n)q^n}
where $\sigma_3(n)$ is the sum of 3th powers of the divisors of $n$.

We discuss briefly how such an identity arises in \vk. One begins
with a comparison of a certain
compactification of the heterotic superstring along $K^3\times T^2$
with a type IIA superstring compactified along a Calabi-Yau
variety $X$. The variety is a degree 12 hypersurface in weighted projective
space $\P^4[1,1,2,2,6]$.
(see \vk.) This hypersurface is singled out
as a candidate on the basis of
matching of the {\it physical spectrum} of the heterotic
theory with the {\it Hodge numbers} of the Calabi-Yau variety on the type II
side.
The two theories are conjectured to
be physically equivalent. As a nontrivial check, some quantum couplings of the
two theories were compared in a suitable boundary component
in the moduli space of $X$. On the heterotic side, an unnormalized coupling is
given by
(see \ref\dWKLL{B.\ de Wit, V.\ Kaplunovsky, J.\ Louis and D.\
L\"{u}st, {\it Perturbative Couplings of Vector Multiplets in $N=2$
Heterotic String Vacua}, hep-th/9504006.}
\ref\AFGNT{I.\ Antoniadis, S.\ Ferrara,
E.\ Gava, K.S.\ Narain and T.R.\ Taylor, {\it Perturbative
Prepotential and Monodromies in N=2 Heterotic Superstring},
hep-th/9504034.})
\eqn\dumb{
F_{\tau\tau\tau}={j'(\tau)^3\over j(\tau)(j(\tau)-j(i))^2}}
where $\tau$ is the period ratio of the $T^2$ on the heterotic side.
The corresponding normalized coupling $\tilde{F}_{\tau\tau\tau}$ differs by an
overall gauge -- a holomorphic function on the upper half plane.
On physical grounds, it is argued that $F_{\tau\tau\tau}/E_4$ should be
equal to the quantum coupling
$K_{t_1 t_1 t_1}$ on $X$ in a
limit when one of the K\"ahler moduli $t_2$ goes to $\infty$, and that $t_1$
should be identified with $\tau$.

Mirror symmetry for $X$ predicts that the normalized coupling is given by
\footnote{$^2$}{Actually this is only one of two terms. The second term is
not relevant in this discussion.}
\eqn\Kttt{
K_{\tau\tau\tau}={1\over w_0(x)^2}({dx\over d\tau})^3 {1\over x^3(1-1728x)^2} }
(see \ref\HKTYI{S.\ Hosono, A.\ Klemm, S.\ Theisen and S.-T. Yau,
Comm. Math. Phys. 167(1995) 301, hep-th/9308122.}
\ref\COFKM{P.\ Candelas, X.\ de la Ossa, A.\ Font, S.\ Katz and D.\
Morrison, Nucl. Phys. 416 (1994) 481, hep-th/9308083.}) where $x=x(\tau)$
is the mirror map restricted along $t_2=\infty$, and $w_0$ is
the holomorphic period for the mirror manifold of $X$.
It has been observed numerically in \COFKM that $x(\tau)={1\over
j(\tau)}$.\lref\lkm{A. Klemm, W. Lerche and P. Mayr, {\sl K3 fibrations and
heterotic type II string duality}, hep-th/9506112.}
\lref\ly{B. Lian and S.-T. Yau, {\sl Arithmetic Properties
of Mirror Maps and Quantum Couplings}, hep-th/9411234.}
Thus the fact that $F_{\tau\tau\tau}/E_4$ should agree with $K_{\tau\tau\tau}$
suggests the identity $w_0(x(\tau))^2=E_4(\tau)$. This has been independently
observed by the authors of \ref\lkt{V. Kaplunovsky, J. Louis and S. Theisen,
{\sl Aspects of duality in N=2 string vacua}, hep-th/9506110.} and by us.
Computing $w_0(x)$, one gets the LHS of \one.

Note that the right hand side \one
 is a modular form of weight 4 while the left hand
side seems to have weight 0!
But the catch is that the left hand side has only
a finite radius of convergence. Thus \one holds only in
the common domain of definitions of both sides, not the whole upper half plane.

Eqn. \one~ is an immediate consequence of two classical identities.
Namely,
\eqn\FF{\eqalign{ {}_2F_1(1/12,5/12; 1; 1728/j)^4=&E_4\cr
{}_2F_1(1/12,5/12;1;z)^2=& {}_3F_2((1/6,5/6,1/2; 1,1; z), }}
the right hand side of the second equation being equal to the left hand side
of \one with $z=1728/j$. The first identity in \FF was already known to
Fricke in his work on elliptic functions \ref\fricke{R. Fricke,
{\sl Elliptischen Funktionen Und Ihre Anwendungen, 1916.}, Chap 6.}.

\newsec{Generalizations}

It is easy to generalize \one to relations involving modular forms for
triangle groups.
Recall that the hypergeometric equation:
\eqn\hyp{z(1-z)y''+(1-{3z\over2})y'-{1\over16}(1-4\nu^2)y=0}
has a unique holomorphic solution $y_0$ near $z=0$ with $y_0(0)=1$, and
a solution $y_1$ with $y_1=y_0Log\ z+O(z)$. The inverse $z(q)$ of the power
series $q=exp\left({y_1\over y_0}\right)=z+O(z^2)$ defines an invertible
holomorphic function in a disc.  The inverse is then denoted by $z(q)$ and
we write $x(q)={1\over\lambda}z(\lambda q)$ for a given $\lambda$.

\proposition{For any complex numbers $\lambda, \nu$ with $\lambda\neq0$, we
have
\eqn\FE{{}_3F_2({1\over2},{1\over2}+\nu,{1\over2}-\nu;1,1;\lambda x(q))^2
={{x'}^2\over x^2(1-\lambda x)}.}
}

Note that $\lambda$ is just a trivial rescaling on both sides of \FE.
Equation \one is the special case with $(\lambda,\nu)=(2^6 3^3, 3^{-1})$,
where the right hand side is the modular form $E_4$ for the group
$\Gamma_0(1)$.
For $(\lambda,\nu)=(2^8,2^{-2}), (2^2 3^3, 2^{-1}3^{-1}), (2^6,0)$, one gets
respectively
\eqn\xiE{\eqalign{
\left(\sum_{n=0}^\infty{(4n)!\over n!^4} x_2(q)^n\right)^2=&
{{x_2'}^2\over x_2^2(1- 256x_2)}\cr
\left(\sum_{n=0}^\infty{(2n)!(3n)!\over n!^5} x_3(q)^n\right)^2=&
{{x_3'}^2\over x_3^2(1-108 x_3)}\cr
\left(\sum_{n=0}^\infty{(2n)!^3\over n!^6} x_4(q)^n\right)^2=&
{{x_4'}^2\over x_4^2(1-64 x_4)}}}
where the right hand sides are weight 4 modular forms of the genus
zero groups
$\Gamma_0(2)+, \Gamma_0(3)+, \Gamma_0(4)+$ respectively
(see \ref\CN{J.H. Conway and S.P. Norton,``Monstrous Moonshine'', Bull. London
Math. Soc., 11(1979) 308-339.} on notations), and $x_2, x_3, x_4$
are their respective
hauptmoduls. It can be shown that \xiE follows from relations of hypergeometric
series similar to \FF. We shall generalize the above identities to
cases involving a class of differential equations of Fuchsian type.
In some special cases, it gives relations involving modular forms for
other genus zero groups which are nontriangle, hence generalizing the above
identities. We now give an elementary proof of the proposition above.

\proof From classical theory of Schwarzian equation, it follows that
$x(q)$ is a solution to
\eqn\Sch{2Qx'^2+\{x,t\}=0}
where
\eqn\dumb{Q={1+(-{5\over4}+\nu^2)\lambda x +(1-\nu)(1+\nu)\lambda^2 x^2\over
4 x^2 (1-\lambda x)^2}. }
On the one hand, ${}_3F_2({1\over2},{1\over2}+\nu,
{1\over2}-\nu;1,1;\lambda x)$ is the unique power series solution
with leading term $1+O(x)$ to the
differential operator ($\Theta_x=x{d\over dx}$):
\eqn\dumb{L=\Theta_x^3-\lambda x(\Theta_x+1/2)(\Theta_x+1/2+\nu)
(\Theta_x+1/2-\nu).}
On the other hand
\eqn\Li{L {x'\over x(1-\lambda x)^{1/2}}
=(1-\lambda x)^{1/2}x^2{x'}^{-2}{d\over dt}(2Q{x'}^2+\{x,t\}).}
 The right hand side vanishes by virtue of the Schwarzian equation.
Now \FE follows from uniqueness. \qed

\subsec{remarks}

It is easy to show that each of the above hauptmoduls is algebraic over
$\Q(j)$.
The explicit polynomial relation between each hauptmodul and $j$ has been
constructed in \ly. Such a polynomial relation, together with each
relation in \xiE, gives yet another relation between a generalized
hypergeometric function $F$, the $j$ function and its derivative.

On the physics side, the above examples also arise
in certain degeneration of Calabi-Yau compactification of type II strings --
in much the same way \one arises in degenerating
a family of degree 12
hypersurfaces in $\P^4[1,1,2,2,6]$ \vk.
The three cases with $(\lambda,\nu)=(2^8,2^{-2}), (2^2 3^3, 2^{-1}3^{-1}),
(2^6,0)$
above correspond respectively to the following
three types of Calabi-Yau varieties in weighted projective spaces:
$X_8(1,1,2,2,2)$, $X_{6,4}(1,1,2,2,2,2)$,
$X_{4,4,4}(1,1,2,2,2,2,2)$ (the list of integers being the weights, the
subscripts being the degrees). These examples have recently been studied
 in \lkm in the context of string duality. We expect the above modular
relations
will be relevant for understanding heterotic-type II duality;
see \vafawitten.

\newsec{Further Generalizations}

In the interesting cases, the
 modular relations derived above involved three ingredients: (1) a modular
function $x$ of a suitable type, (2) a modular form $E$ of weight 4,
and (3) a power series solution $w_0$
 of a third order differential operator $L$. The $x$
satisfies a Schwarzian equation \Sch determined by a rational function $Q$.
The $E$ is an algebraic expression
of $x,x'$. And the monodromy of solutions near $x=0$ to
operator $L$ has maximal unipotency. (A consequence of this is the
uniqueness of power series solution $w_0$.)

Given a genus zero group $G$
and a hauptmodul $x(q)$, it's easy to show that
the Schwarzian derivative
\eqn\dumb{\{x,t\}={x'''\over x'}-{3\over 2}({x''\over x'})^2 }
is a modular form of weight 4. Thus it takes the form $-2Q{x'}^2$ (the -2 is
just for convenience) for some rational function $Q$. To obtained a relation
analogous to \FE, our proof suggests that we should construct a
weight 4 modular form $E=x'^2/r(x)$ and an operator
$L$ whose monodromy has maximal unipotency at $x=0$, such that
$LE^{1/2}=0$. It turns out that in Mirror Symmetry,
there is an abundance of operators with maximal unipotent monodromy.
The four explicit examples we have seen come in fact from the following
1-parameter families of K3 surfaces in weighted projective spaces:
$X_6(1,1,1,3)$, $X_4(1,1,1,1)$, $X_{2,3}(1,1,1,1,1)$, $X_{2,2,2}(1,1,1,1,1,1)$.
The corresponding third order differential operators $L$ in these cases are
precisely the Picard-Fuchs operators. The same
operators also arise from certain degenerations of Calabi-Yau threefolds;
see \lkm and \vafawitten.

In the following we construct a list of examples using mirror symmetry.
Each of the genus zero groups and its hauptmodul are
 of the types considered in \CN. The
corresponding  rational function $Q$ will be given. We write it in the form
$Q={p\over 4r^2}$ where $p,r$ are polynomials which are relatively prime.
The weight 4 modular form
takes the form ${x'^2\over x r(x)}$. The operator $L$ will be the Picard-Fuchs
operator obtained by
degenerating a given family of K3 surfaces
which are complete intersections of $l$ hypersurfaces in a product of $k$
projective spaces (see \ref\hktyI{S. Hosono, A. Klemm, S. Theisen and
S.-T. Yau, Comm. Math Phys. 167 (1995) 301-350.} and references therein.)
The degrees of the $l$ hypersurfaces are given by
the matrix ${\bf d}=(d^{(i)}_j)_{i=1,..,k;j=1,..,l}$. The dimensions of
the $i$th projective space is denoted by $D_i$. Note that the adjunction
formula requires that for each $i$,
\eqn\dumb{D_i+1=\sum_{j=1}^l d^{(i)}_j.}
The deformation parameters of this family are denoted by
$x_1,..,x_k$ as defined in \hktyI.
The following table gives a list of examples up to $k\leq4$.)
The degeneration is along the ``diagonal'' $x:=x_1=\cdots=x_k$.
The $L$ is an operator of Fuchsian type and is of the form
\eqn\PFL{L=\Theta_x^3 - \sum_{i=1}^m \lambda_i x^i p_i(\Theta_x)}
where the $p_i$ are monic polynomials of degree 3. It is easy to see
that $L$ has a unique power series solution $w_0$ with $w_0(0)=1$. In fact its
coefficients are determined by the recursion relation:
\eqn\dumb{A_n={1\over n^3}\sum_{i=1}^mp_i(n-i)A_{n-i}.}
There is also a standard construction for $w_0(x)$ in mirror symmetry. It can
be obtained by restricting the holomorphic period of the $k$-moduli
K3 surfaces along the diagonal  $x:=x_1=\cdots=x_k$. The general formula
is given by
\eqn\dumb{
w_0(x)=\sum_{n_1+\cdots+n_k=n} {\Pi_{j=1}^l(n_1 d^{(1)}_j+\cdots+n_k
d^{(k)}_j)!
\over \Pi_{i=1}^k n_i!^{D_i+1} } x^n }
where the sum is over all nonnegative $n_1,..,n_k,n$.

The modular relation we obtain takes the form (cf. \FE)
\eqn\dumb{w_0(x)^2={x'^2\over x r(x)}}
It can be proved in a completely analogous ways as for \FE.
First we check that
\eqn\dumb{L{x'\over (x r(x))^{1/2}}\equiv0~~mod~~ 2Qx'^2+\{x,t\}.}
Now use the fact that $x(t)$ a hauptmodul of the genus zero group $G$
\footnote{${}^\dagger$}{See \CN on notations.},
and hence satisfies a Schwarzian equation given by $2Qx'^2+\{x,t\}=0$.

We have also included three examples (end of table below) for which
we know a modular relation and the genus zero group, but we do not
know the corresponding family of K3 surfaces, or if it exists.

\vfill\eject

$${\fivepoint{
\vbox{\offinterlineskip\tabskip=0pt
\halign{\strut\vrule#
&\hfil~$#$
&~~$#$~~\hfil
&~~$#$~~\hfil
&~~$#$~~\hfil
&\vrule#\cr
\noalign{\hrule}
& {\rm degree~ matrices~}{\bf d}&
{\rm  diff.~ operators~}L& {\rm potential~}Q(x) & {\rm genus~0~groups~}G
&\cr
\noalign{\hrule}
&
\left(\matrix{
 4
}\right)
&
\Theta_x^3 - 8x(1 + 2 \Theta_x)(1 + 4 \Theta_x)(3 + 4 \Theta_x)
&
{1 - 304x + 61440x^2 \over 4(1 - 256x)^2x^2 }
&
\Gamma_0(2)+ &\cr
\noalign{\hrule}
&
\left(\matrix{
2& 3
}\right)
&
\Theta_x^3 - 6x(1 + 2 \Theta_x)(1 + 3 \Theta_x)(2 + 3 \Theta_x)
&
{1 - 132x + 11340x^2 \over 4(1 - 108x)^2x^2 }
&
\Gamma_0(3)+ &\cr
\noalign{\hrule}
&
\left(\matrix{
2& 2& 2
}\right)
&
\Theta_x^3 - 8x(1 + 2 \Theta_x)^3
&
{1 - 80x + 4096x^2 \over 4(1 - 64x)^2x^2 }
&
\Gamma_0(4)+ &\cr
\noalign{\hrule}
&
\left(\matrix{
2\cr 3
}\right)
&
\Theta_x^3 - 8x(1 + 2 \Theta_x)^3
&
{1 - 80x + 4096x^2 \over 4(1 - 64x)^2x^2 }
&
\Gamma_0(4)+ &\cr
\noalign{\hrule}
&
\left(\matrix{
0& 2\cr 2& 2
}\right)
&
\eqalign{\Theta_x^3 +& 36x^2(1 +  \Theta_x)(1 + 2 \Theta_x)(3 + 2 \Theta_x)\cr
-& 2x(1 + 2 \Theta_x)(3 + 10 \Theta_x + 10 \Theta_x^2)}
&
{1 - 52x + 1500x^2 - 6048x^3 + 15552x^4 \over
   4(1 - 36x)^2(1 - 4x)^2x^2 }
&
\Gamma_0(6)+ &\cr
\noalign{\hrule}
&
\left(\matrix{
1& 1\cr 2& 2
}\right)
&
\eqalign{\Theta_x^3 +& x^2(1 +  \Theta_x)^3 \cr
- &   x(1 + 2 \Theta_x)(5 + 17 \Theta_x + 17 \Theta_x^2)}
&
{1 - 44x + 1206x^2 - 44x^3 + x^4 \over 4x^2(1 - 34x + x^2)^2 }
&
\Gamma_0(6)+6 &\cr
\noalign{\hrule}
&
\left(\matrix{
1& 2\cr 1& 2
}\right)
&
\eqalign{\Theta_x^3 - &32x^2(1 +  \Theta_x)(1 + 2 \Theta_x)(3 + 2 \Theta_x) \cr
- &   2x(1 + 2 \Theta_x)(2 + 7 \Theta_x + 7 \Theta_x^2)}
&
{1 - 36x + 972x^2 + 3712x^3 + 12288x^4 \over
   4(-1 - 4x)^2x^2(-1 + 32x)^2 }
&
\Gamma_0(6)+ &\cr
\noalign{\hrule}
&
\left(\matrix{
1& 2\cr 2& 1
}\right)
&
\eqalign{\Theta_x^3 -& 3x^2(1 +  \Theta_x)(2 + 3 \Theta_x)(4 + 3 \Theta_x) \cr
-&  x(1 + 2 \Theta_x)(4 + 13 \Theta_x + 13 \Theta_x^2)}
&
{1 - 34x + 745x^2 + 840x^3 + 648x^4 \over
   4(-1 - x)^2x^2(-1 + 27x)^2 }
&
\Gamma_0(7)+ &\cr
\noalign{\hrule}
&
\left(\matrix{
0& 1& 1& 2\cr 2& 1& 1& 0
}\right)
&
\eqalign{\Theta_x^3 +& 64x^2(1 +  \Theta_x)^3 \cr
- &  2x(1 + 2 \Theta_x)(2 + 5 \Theta_x + 5 \Theta_x^2)}
&
{1 - 28x + 396x^2 - 1792x^3 + 4096x^4 \over
   4(1 - 16x)^2(1 - 4x)^2x^2 }
&
\Gamma_0(12)+ &\cr
\noalign{\hrule}
&
\left(\matrix{
1& 1& 1& 1\cr 1& 1& 1& 1
}\right)
&
\eqalign{\Theta_x^3 -& 4x^2(1 +  \Theta_x)(3 + 4 \Theta_x)(5 + 4 \Theta_x)\cr
 - &   2x(1 + 2 \Theta_x)(1 + 3 \Theta_x + 3 \Theta_x^2)}
&
{1 - 16x + 224x^2 + 976x^3 + 3840x^4 \over
   4(-1 - 4x)^2x^2(-1 + 16x)^2 }
&
\Gamma_0(10)+ &\cr
\noalign{\hrule}
&
\left(\matrix{
2\cr 2\cr 2
}\right)
&
\eqalign{\Theta_x^3 +& 36x^2(1 +  \Theta_x)(1 + 2 \Theta_x)(3 + 2 \Theta_x)\cr
 - &   2x(1 + 2 \Theta_x)(3 + 10 \Theta_x + 10 \Theta_x^2)}
&
{1 - 52x + 1500x^2 - 6048x^3 + 15552x^4 \over
   4(1 - 36x)^2(1 - 4x)^2x^2 }
&
\Gamma_0(6)+ &\cr
\noalign{\hrule}
&
\left(\matrix{
1& 1& 0\cr 1& 1& 0\cr 1& 1& 2
}\right)
&
\eqalign{\Theta_x^3 + &64x^2(1 +  \Theta_x)^3 \cr
- &   2x(1 + 2 \Theta_x)(2 + 5 \Theta_x + 5 \Theta_x^2)}
&
{1 - 28x + 396x^2 - 1792x^3 + 4096x^4 \over
   4(1 - 16x)^2(1 - 4x)^2x^2 }
&
\Gamma_0(12)+ &\cr
\noalign{\hrule}
&
\left(\matrix{
0& 0& 2\cr 1& 1& 1\cr 1& 1& 1
}\right)
&
\eqalign{\Theta_x^3 + &64x^2(1 +  \Theta_x)^3 \cr
- &   2x(1 + 2 \Theta_x)(2 + 5 \Theta_x + 5 \Theta_x^2)}
&
{1 - 28x + 396x^2 - 1792x^3 + 4096x^4 \over
   4(1 - 16x)^2(1 - 4x)^2x^2 }
&
\Gamma_0(12)+ &\cr
\noalign{\hrule}
&
\left(\matrix{
1& 0& 1\cr 0& 1& 2\cr 2& 1& 0
}\right)
&
\eqalign{\Theta_x^3 -& 98x^3(1 +  \Theta_x)(2 +  \Theta_x)(3 + 2 \Theta_x) \cr
- &   x(1 + 2 \Theta_x)(5 + 11 \Theta_x + 11 \Theta_x^2) \cr
+ &   x^2(1 +  \Theta_x)(141 + 242 \Theta_x + 121 \Theta_x^2)}
&
{1 - 32x + 482x^2 - 3332x^3 + 12553x^4 - 27636x^5 +
       28812x^6 \over 4x^2(-1 + 4x)^2(-1 + 18x - 49x^2)^2 }
&
\Gamma_0(14)+ &\cr
\noalign{\hrule}
&
\left(\matrix{
0& 0& 1& 2\cr 0& 2& 1& 0\cr 2& 0& 1& 0
}\right)
&
\eqalign{\Theta_x^3 - &192x^3(1 +  \Theta_x)(2 +  \Theta_x)(3 + 2 \Theta_x) \cr
- &   6x(1 + 2 \Theta_x)(1 + 2 \Theta_x + 2 \Theta_x^2)\cr
 + &   4x^2(1 +  \Theta_x)(51 + 88 \Theta_x + 44 \Theta_x^2)}
&
{1 - 36x + 572x^2 - 5088x^3 + 26688x^4 - 78336x^5 +
       110592x^6 \over 4(1 - 8x)^2x^2(-1 + 4x)^2(-1 + 12x)^2 }
&
\Gamma_0(12|2)+ &\cr
\noalign{\hrule}
}
\hrule}
}}$$

$${\fivepoint{
\vbox{\offinterlineskip\tabskip=0pt
\halign{\strut\vrule#
&\hfil~$#$
&~~$#$~~\hfil
&~~$#$~~\hfil
&~~$#$~~\hfil
&\vrule#\cr
\noalign{\hrule}
& {\rm degree~ matrices~}{\bf d}&
{\rm  diff.~ operators~}L& {\rm potential~}Q(x) & {\rm genus~0~groups~}G &\cr
\noalign{\hrule}
&
\left(\matrix{
0& 1& 1& 1\cr 1& 0& 1& 1\cr 1& 1& 0& 1
}\right)
&
\eqalign{\Theta_x^3 -& 30x^3(1 +  \Theta_x)(2 +  \Theta_x)(3 + 2 \Theta_x) \cr
- &   x(1 + 2 \Theta_x)(3 + 7 \Theta_x + 7 \Theta_x^2)\cr
 + &   x^2(1 +  \Theta_x)(33 + 58 \Theta_x + 29 \Theta_x^2)}
&
{1 - 20x + 206x^2 - 336x^3 + 57x^4 - 1980x^5 + 2700x^6 \over
   4x^2(-1 + 12x)^2(-1 + 2x - 5x^2)^2 }
&
\Gamma_0(15)+ &\cr
\noalign{\hrule}
&
\left(\matrix{
1& 1& 0& 1& 0\cr 1& 1& 1& 0& 0\cr 0& 0& 1& 1& 2
}\right)
&
\eqalign{\Theta_x^3 - &8x^3(3 + 2 \Theta_x)^3\cr
- &   4x(1 + 2 \Theta_x)(1 + 2 \Theta_x + 2 \Theta_x^2)\cr
+ &   16x^2(1 +  \Theta_x)(5 + 8 \Theta_x + 4 \Theta_x^2)}
&
{1 - 24x + 272x^2 - 1488x^3 + 4352x^4 - 6144x^5 + 4096x^6 \over
   4x^2(-1 + 4x)^2(-1 + 12x - 16x^2)^2 }
&
\Gamma_0(20)+ &\cr
\noalign{\hrule}
&
\left(\matrix{
1& 1\cr 1& 1\cr 1& 1\cr 1& 1
}\right)
&
\eqalign{\Theta_x^3 +& 64x^2(1 +  \Theta_x)^3 \cr
- &   2x(1 + 2 \Theta_x)(2 + 5 \Theta_x + 5 \Theta_x^2)}
&
{1 - 28x + 396x^2 - 1792x^3 + 4096x^4 \over
   4(1 - 16x)^2(1 - 4x)^2x^2 }
&
\Gamma_0(12)+ &\cr
\noalign{\hrule}
&
\left(\matrix{
0& 1& 1& 0\cr 0& 1& 1& 0\cr 0& 0& 1& 2\cr 2& 0& 1& 0
}\right)
&
\eqalign{\Theta_x^3 - &192x^3(1 +  \Theta_x)(2 +  \Theta_x)(3 + 2 \Theta_x) \cr
- &   6x(1 + 2 \Theta_x)(1 + 2 \Theta_x + 2 \Theta_x^2)\cr
 + &   4x^2(1 +  \Theta_x)(51 + 88 \Theta_x + 44 \Theta_x^2)}
&
{1 - 36x + 572x^2 - 5088x^3 + 26688x^4 - 78336x^5 +
       110592x^6 \over 4(1 - 8x)^2x^2(-1 + 4x)^2(-1 + 12x)^2 }
&
\Gamma_0(12|2)+ &\cr
\noalign{\hrule}
&
\left(\matrix{
1& 0& 0& 1\cr 1& 0& 0& 1\cr 0& 1& 1& 1\cr 1& 1& 1& 0
}\right)
&
\eqalign{\Theta_x^3 - &8x^3(3 + 2 \Theta_x)^3\cr
 - &   4x(1 + 2 \Theta_x)(1 + 2 \Theta_x + 2 \Theta_x^2)\cr
 + &   16x^2(1 +  \Theta_x)(5 + 8 \Theta_x + 4 \Theta_x^2)}
&
{1 - 24x + 272x^2 - 1488x^3 + 4352x^4 - 6144x^5 + 4096x^6 \over
   4x^2(-1 + 4x)^2(-1 + 12x - 16x^2)^2 }
&
\Gamma_0(20)+ &\cr
\noalign{\hrule}
&
\left(\matrix{
0& 0& 0& 1& 1& 1\cr 0& 0& 1& 0& 1& 1\cr
   1& 1& 0& 1& 0& 0\cr 1& 1& 1& 0& 0& 0
}\right)
&
\eqalign{\Theta_x^3 - &14x^3(1 +  \Theta_x)(2 +  \Theta_x)(3 + 2 \Theta_x) \cr
- &   24x^4(2 +  \Theta_x)(3 + 2 \Theta_x)(5 + 2 \Theta_x)\cr
- &   x(1 + 2 \Theta_x)(4 + 7 \Theta_x + 7 \Theta_x^2)\cr
 + &   x^2(1 +  \Theta_x)(72 + 106 \Theta_x + 53 \Theta_x^2)}
&
\eqalign{(1 -& 22x + 225x^2 - 1292x^3 + 4436x^4 \cr-& 8304x^5 +
       5124x^6 + 2016x^7 + 6912x^8)/ \cr
   & (4(1 - 8x)^2(1 - 4x)^2(-1 - x)^2x^2(-1 + 3x)^2) }
&
\Gamma_0(30)+ &\cr
\noalign{\hrule}
&
\left(\matrix{
0& 0& 0& 1& 1& 1\cr 0& 1& 1& 0& 0& 1\cr
   1& 0& 1& 0& 1& 0\cr 1& 1& 0& 1& 0& 0
}\right)
&
\eqalign{\Theta_x^3 +& 64x^4(2 +  \Theta_x)^3\cr
 - &   x(1 + 2 \Theta_x)(4 + 7 \Theta_x + 7 \Theta_x^2)\cr
- &   8x^3(3 + 2 \Theta_x)(18 + 21 \Theta_x + 7 \Theta_x^2)\cr
 + &   x^2(1 +  \Theta_x)(88 + 122 \Theta_x + 61 \Theta_x^2)}
&
\eqalign{(1 -& 22x + 225x^2 - 1224x^3 + 4168x^4 \cr- &9792x^5 +
       14400x^6 -11264x^7 + 4096x^8)/ \cr
   &(4(1 - 8x)^2(1 - x)^2x^2(1 - 5x + 8x^2)^2 )}
&
\Gamma_0(28)+ &\cr
\noalign{\hrule}
&
\left(\matrix{??
}\right)
&
\eqalign{\Theta_x^3 - &  54x(1 + 2 \Theta_x)(4 + 9 \Theta_x + 9 \Theta_x^2)\cr
 + & 81  x^2(1 +  \Theta_x)(2+3\Theta_x)(4+3\Theta_x)}
&
{1 - 24 x + 648 x^2\over 4 x^2  (-1 + 27 x)^2}
&
\Gamma_0(2)- &\cr
\noalign{\hrule}
&
\left(\matrix{??
}\right)
&
\eqalign{\Theta_x^3 - &  8x(1 + 2 \Theta_x)(3 + 8 \Theta_x + 8 \Theta_x^2)\cr
 + & 1024  x^2(1 +  \Theta_x)(1+2\Theta_x)(3+2\Theta_x)}
&
{1 -48 x + 3072 x^2\over 4 x^2  (-1 + 64 x)^2}
&
\Gamma_0(3)- &\cr
\noalign{\hrule}
&
\left(\matrix{??
}\right)
&
\eqalign{\Theta_x^3 - &  8x(1 + 2 \Theta_x)(1 + 2 \Theta_x + 2 \Theta_x^2)\cr
 + & 256 x^2(1 +  \Theta_x)^3}
&
{1 -16 x + 256 x^2\over 4 x^2  (-1 + 16 x)^2}
&
\Gamma_0(3)- &\cr
\noalign{\hrule}
}
\hrule}
}}$$

Two observations: 1. in each case above, the
differential operator $L$ always takes the form
\eqn\genL{
L=\Theta_x^3-\sum_{i=1}^m\lambda_ix^i(\Theta_x+i/2)(\Theta_x+i/2+\nu_i)
(\Theta_x+i/2-\nu_i)}
where the $\lambda_i,\nu_i$ are algebraic numbers.
We denote the power series solution to $L$ with leading coefficient 1 as
$w(\lambda;\nu;x)$.
2. The corresponding
 hauptmodul $x(q)$ is a solution to a Schwarzian equation \Sch,
hence $x(q)$ is the function defined by the inverse of the relation
$q=exp\left(y_1(x)/y_0(x)\right)$ where $y_1,y_0$ are suitable solutions
to the ODE $y''+Qy=0$. (When the modular group is triangle, $x(q)$ is
nothing but the inverse of a triangle function.)

Given $\lambda=(\lambda_1,..,\lambda_m)$, $\nu=(\nu_1,..,\nu_m)$ as in \genL,
consider the differential operator
\eqn\Ltilde{\tilde{L}=\Theta_x^2-\sum_{i=1}^m\lambda_ix^i(\Theta_x+i/4+\nu_i/2)
(\Theta_x+i/4-\nu_i/2).}
The singularities of \Ltilde are $\infty,0$, which are
both regular, and the roots of
$1-\lambda_1x-\cdots-\lambda_mx^m=0$, each of which is regular if its
multiplicity does not exceed 2. In this case, \Ltilde is Fuchsian.

The operator $\tilde{L}$ has a power series solution
\eqn\dumb{\tilde{w}(\lambda;\nu;x)=\sum_{n=0}^\infty A_n x^n}
determined by $A_0=1$ and the recursion relation
\eqn\dumb{A_n={1\over n^2}\sum_{i=1}^m\lambda_i (n-i+i/4+\nu_i/2)
(n-i+i/4-\nu_i/2)A_{n-i}.}
A second solution takes the form $w(\lambda;\nu;x)Log\ x +g(x)$ where
$g(x)$ is a power series determined by $g(0)=0$ and the relation
\eqn\dumb{\tilde{L}g=-2\Theta_x w+ 2\sum_{i=1}^m\lambda_i x^i (\Theta_x+i/4)w.}
Thus we define
\eqn\dumb{x(\lambda;\nu;q)=x(q)=inverse~of~the~relation~q=x~
exp\left(g/\tilde{w}\right).}

\theorem{Given $\lambda,\nu$ as above, we have the identity
\eqn\dumb{w(\lambda;\nu;x)^2={x'^2\over x^2(1-\lambda_1x-\cdots-\lambda_mx^m)}
}
}
 This subsumes all the examples given above.

\proof We will prove the following identities
\eqn\steps{\eqalign{
1.~& w(\lambda;\nu;x)=\tilde{w}(\lambda;\nu;x)^2\cr
2.~ &  \tilde{w}(\lambda;\nu;x)^4={x'^2\over
x^2(1-\lambda_1x-\cdots-\lambda_mx^m)} } }
from which our theorem follows immediately.

To prove 1.,  by the uniqueness of power series solution to $L$, it is enough
to prove that $L\tilde{w}^2=0$. Using $\tilde{L}\tilde{w}=0$, we do
\eqn\dumb{\eqalign{
\Theta_x^3 \tilde{w}^2=&2\tilde{w}\Theta_x^3\tilde{w}+
6\Theta_x\tilde{w}\Theta_x^2\tilde{w}\cr
=&\sum_{i=1}^m (2\tilde{w}\Theta_x+6\Theta_x\tilde{w})
\lambda_ix^i(\Theta_x+i/4+\nu_i/2)(\Theta_x+i/4-\nu_i/2)\tilde{w}\cr
=&\sum_{i=1}^m \lambda_ix^i (2\tilde{w}(\Theta_x+i)+6\Theta_x\tilde{w})
(\Theta_x+i/4+\nu_i/2)(\Theta_x+i/4-\nu_i/2)\tilde{w}. }}
Combining this with the identity
\eqn\dumb{\eqalign{
(2f(\Theta_x+i)+6\Theta_xf)& (\Theta_x+i/4+\nu_1/2)(\Theta_x+i/4-\nu_i/2)f\cr
=&(\Theta_x+i/2)(\Theta_x+i/2+\nu_i)(\Theta_x+i/2-\nu_i)f^2}}
it follows immediately that $L\tilde{w}^2=0$.

To prove 2., by the uniqueness of power series solution to $\tilde{L}$, it is
enough to prove that $\tilde{L}{x'^{1/2}\over
x^{1/2}(1-\lambda_1x-\cdots-\lambda_mx^m)^{1/4}}=0$. Write
$A=x^{-1/2}(1-\lambda_1x-\cdots-\lambda_mx^m)^{-1/4}$.
 Rewriting $\tilde{L}f$,
we get
\eqn\tildeLf{\tilde{L}f=x^2(1-\sum\lambda_ix^i){d^2f\over dx^2}
+x(1-\sum\lambda_ix^i(1+i/2)){df\over dx}
-\sum\lambda_i x^i(i^2/16-\nu_i^2/4)f.}
Recall that under the change of variable
$f=exp\left(-\int{q\over 2p}\right)g$,
the second order expression $p{d^2f\over dx^2}+q{df\over dx}+rf$
goes into its reduced form ${d^2g\over dx^2}+Qg$ for some $Q$.
In our case \tildeLf, the change of variable is simply
$f=x^{-1/2}(1-\lambda_1x-\cdots-\lambda_mx^m)^{-1/4}g=Ag$.
Since $x(q)$ is the inverse of the ratio of two solutions $Ag_0,Ag_1$
 to $\tilde{L}$,
its easy to see that $x'=g_0^2/Wronskian(g_0,g_1)$. Since the wronskian
is constant it follows that $x'^{1/2}$ is proportional to $g_0$.
Thus $\tilde{L}{x'^{1/2}\over x^{1/2}(1-\lambda_1x-\cdots-\lambda_mx^m)^{1/4}}$
is proportional to
$\tilde{L}Ag_0$, which is zero.
This proves our claim. \qed

A simple computation shows that
the reduced form  ${d^2\over dx^2}+Q(x)$ of \Ltilde has
\eqn\Qexp{Q={1\over 4p_2^2}(-p_1^2+4p_0 p_2 - 2p_2 p_1' + 2p_1 p_2')}
where the $p_i$ is the coefficient of ${d^if\over dx^i}$ in \tildeLf.

We end this section with two other observations. The first gives the
reduced form of our operator $L$. The second is an analogue of formula 2
above, but for Gauss' {\it hypergeometric series}.

\proposition{The operator $L$ \genL has the reduced form
${d^3\over dx^3}+4Q(x){d\over dx}+2Q'(x)$ where $Q$ is given by \Qexp. }

\proof A direct computation of the change of variable seems too tedious.
We will thus give a more conceptual proof. Given a general third
order operator $L=\sum_{i=0}^3 p_i(x){d^i\over dx^i}$, its reduced form
becomes ${d^3\over dx^3}+ 4Q(x){d\over dx}+2Q'(x)$ for some $Q$ iff the
following differential equation holds:
\eqn\Qcond{\eqalign{
&-4p_2^3 + 18 p_1 p_2 p_3-54p_0 p_3^2 +27p_3^2 p_1'
-18p_2 p_3 p_2'+18p_2^2 p_3' \cr
&-27p_1 p_3 p_3'+ 18 p_3 p_2' p_3' - 18 p_2 {p_3'}^2 -
9 p_3^2 p_2'' + 9 p_2 p_3 p_3'' = 0. }}
It is easy to check that the form of this condition is independent of
coordinate. That is, if we wrote $L$ in a different coordinate $z$:
$L=\sum_{i=0}^3 q_i(z){d^i\over dz^i}$, then its reduced form becomes
${d^3\over dz^3}+ 4R(z){d\over dx}+2R'(z)$ for some $R$ iff \Qcond holds
with the $p$ replaced by the $q$. Moreover, if $L$ has such a reduced
form in one coordinate $z$, it does so in any other coordinate.
Thus to check condition \Qcond, we choose the simplest coordinate.

In our case \genL, we choose $z=Log~x$ (cf. $\Theta_x={d\over dLog~x}$).
Then \genL becomes $L=\sum_{i=0}^3 q_i(z){d^i\over dz^i}$ with
\eqn\dumb{\eqalign{
q_3=&1-\sum\lambda_i e^{iz}\cr
q_2=&-{3\over 2}\sum\lambda_i i e^{iz}\cr
q_1=&-\sum\lambda_i e^{iz}(3i^2/4 -\nu_i^2)\cr
q_0=&-\sum\lambda_i e^{iz}(i^3/8 -i\nu_i^2/2). }}
Observe that we get $q_2={3\over2}q_3'$, $q_1={3\over4}q_3''+r$,
$q_0={1\over8}q_3'''+{1\over2}r'$ where $r(z)=\sum\lambda_i \nu_i^2 e^{iz}$.
Now substitue these relations together with $p_i=q_i$ into \Qcond, we see that
this condition holds
identically. This shows that the reduced form  of $L$ becomes
${d^3\over dx^3}+4Q(x){d\over dx}+2Q'(x)$ for some $Q$.

Now if we write $L=\sum_{i=0}^3 q_i(x){d^i\over dx^i}$, then a simple
computation gives
\eqn\Qq{Q={1\over 12 q_3^2}(-q_2^2+3q_1 q_3 - 3q_3 q_2'+ 3q_2 q_3').}
Also from the explicit form of $L$ we get
\eqn\qrs{\eqalign{
q_3=&x r\cr
q_2=&3x r'/2\cr
q_1=&x r''/2 + 4x s\cr
q_0=&-\sum\lambda_i x^i(i-2\nu_i)(i+2\nu_i)i/8} }
where $r(x):=x^2(1-\sum\lambda_i x^i)$,
$s(x):=-\sum\lambda_ix^i (i^2/16-\nu_i^2/4)$. Now from \Qexp, \tildeLf, we get
\eqn\prs{\eqalign{
p_2=&r\cr
p_1=&r'/2\cr
p_0=&s.}}
Substituting \prs into \Qexp, \qrs into \Qq, we see that the two expressions
coincide. \qed

Consider the differential operator for the hypergeometric equation:
\eqn\gauss{\Theta_x^2-\lambda x(\Theta_x+a)(\Theta_x+b).}
Once again it has a solution $\tilde{w}_0$,
 regular at $x=0$, with leading term 1,
and a solution $\tilde{w}_1$ with leading term $Log~x$. We can define $x(q)$
as before to be the inverse of the power series relation
$q=e^{\tilde{w}_1(x)/\tilde{w}_0(x)}$.

\proposition{Let $x(q)$ be as just defined. Then
\eqn\dumb{{}_2F_1(a,b;1;\lambda x(q))^4=
{x'^2\over x^2(1-\lambda x)^{2(a+b)}}.} }

\proof In the proof of the theorem above, formula 2 covers the case
(when $m=1$) in which $a+b=1/2$. The argument for general $a,b$
is completely analogous. Indeed, we check that
${x'^{1/2}\over x^{1/2}(1-\lambda x)^{(a+b)/2}}$ is a solution to \gauss~
and is regular at $x=0$ with leading term $1$. Thus it must coincide
with ${}_2F_1(a,b;1;\lambda x)$. \qed

\subsec{remarks}

Existence of relations involving power series
solution to second and third order Fuchsian equations
and modular forms,  clearly suggests similar relations involving
series solutions to ODEs of higher order. Remarkably,
there indeed exists such generalizations.
 Namely it is an identity of the form
\eqn\dumb{w(x)^2=x'^{N-1}R(x)}
where $w=1+O(x)$ is a series solution to certain $N$th order
Fuchsian ODE, and $R$ is a suitable algebraic function whose singularities
lie along those of the ODE. Under suitable hypothesis, the right hand side
is a modular form of weight $2N-2$ of an appropriate type.
 This will be discussed in details in a future paper
\ref\lytwo{B. Lian and S.T.-Yau, work in progress}.

\newsec{Part B. Integrality of Mirror Maps}

The set up of the problem is as follows. Recall that the a smooth degree $N$
Calabi-Yau hypersurface $X$ of dimension $N-2$ has as its mirror a canonical
family of toric hypersurfaces $X^*$ with $h^{N-3,1}(X^*)=1$
\ref\gp{B. Greene and R. Plesser, Nucl. Phys. B 338 (1990),15-37.}\cdgp
(see also \ref\emm{Essays on Mirror Manifolds, Ed. S.-T. Yau,
(International Press, Hong Kong 1992)}).
This family fibers over $\P^1-\{0,N^N,\infty\}$. The Picard-Fuchs equation for
this family is given by
\eqn\dumb{\left(\Theta_z^{N-1}-Nz(N\Theta_z+1)\cdots(N\Theta_z+N-1)\right)f(z)=0.}
By the Frobenius method, a basis of solutions is given by
\eqn\dumb{\omega_i(z)=\left({1\over 2\pi
i}{\partial\over\partial\rho}\right)^i\sum_{k\geq0}
{\Gamma(N(k+\rho)+1)\over\Gamma(k+\rho+1)^N}|_{\rho=0}}
$i=0,..,N-1$.

Let $J$ be the Fubini-Studi K\"ahler class pullbacked to $X$, $\cC$ be
the real K\"ahler cone, and $\cK=i\cC+H^2(X,\R)/H^2(X,\Z)$ the complexified
K\"ahler cone of $X$. The mirror map is defined locally as a map from a domain
$\cK(r)=\{tJ|Im~t>r\}\subset\cK$ with $r>>0$, to $\P^1$ regarded as a
deformation space of complex structures for
the mirror family $X^*$. If we let $q=e^{2\pi it}$, then the map is given by
the q-series which the inverse of the relation
\eqn\dumb{q=exp({\omega_1(z)\over\omega_0(z)})}

Our problem is to give a complete proof that all the coefficients in the
q-series $z(q)$ are integral. For simplicity, we will restrict to the case when
$N$ is prime.

\theorem{For each prime $N$, the coefficients of the
q-series $z(q)$ defined above are rational integers.}

\proof By induction, it's easy to see that $z(q)$ is integral iff the inverse
$q(z)$ is integral. Now $q(z)=exp({\omega_1(z)\over\omega_0(z)})=z
e^{{g(z)\over\omega_0(z)}}$ for some $g(z)\in z\Q[[z]]$. We will prove that
$e^{{g(z)\over\omega_0(z)}}\in 1+X\Z_p[[X]]$ for all prime $p$.

To write down $g(z)$ explicitly, let
\eqn\eqnD{\eqalign{
D_x(m)=&\sum_{j=0}^{m-1}{1\over x+j}\cr
D(m)=&\sum_{i=1}^{N-1} D_{{i\over N}}(m)-(N-1)D_1(m).
}}
Then it's easy to check that
\eqn\eqng{g(z)=\sum_{m>0}{(Nm)!\over (m!)^N}D(m) z^m.}

\newsec{Step 1: Dwork's Lemma}
\seclab\partI

\lemma{ Let $F(X)=\sum a_iX^i\in 1+X\Q_p[[X]]$. Then $F(X)\in 1+X\Z_p[[X]]$ iff
$F(X^p)/F(X)^p\in1+pX\Z_p[[X]]$.}

For proof, see \ref\Koblitz{N. Koblitz, {\sl p-adic Numbers, p-adic
Analysis, and Zeta-Functions}, Springer Verlag 1977, p84.}

\corollary{Let $f(X)\in X\Q_p[[X]]$. Then $e^{f(X)}\in1+X\Z_p[[X]]$ iff
$f(X^p)-pf(X)\in pX\Z_p[[X]]$.}

\proof Write $F(X)=e^{f(X)}$ which is in $1+X\Q_p[[X]]$. If
$e^{f(X)}\in1+X\Z_p[[X]]$, then by Dwork's Lemma,
\eqn\dumb{F(X^p)/F(X)^p=e^{f(X^p)-pf(X)}\in 1+pX\Z_p[[X]].}
Write the RHS as $1-pXG(X)$ with $G(X)\in\Z_p[[X]]$. Then
\eqn\dumb{f(X^p)-pf(X)=log_p(1-pXG(X))=\sum_{i>0}{(pXG(X))^i\over i}.}
But $p^i/i\in p\Z_p$ for all $i>0$. Thus $f(X^p)-pf(X)\in pX\Z_p[[X]]$.

Conversely suppose $f(X^p)-pf(X)=pX H(X)$ with $H(X)\in \Z_p[[X]]$. Using the
fact that for $n>0$, $ord_p{p^n\over n!}=n-{n-S(n)\over p-1}>0$ where $S(n)$ is
the sum of the p-adic digits of $n$, we see that
\eqn\dumb{{e^{f(X^p)}\over e^{pf(X)}}=e^{pXH(X)}=1+\sum_{n>0} {p^n\over
n!}X^nH(X)^n\in 1+X\Z_p[[X]].}
Thus by Dwork's Lemma, we conclude that $e^{f(X)}\in1+X\Z_p[[X]]$. \qed

\bs
If we let
\eqn\dumb{f(z)={g(z)\over\omega_0(z)}={\sum_{m>0}{(Nm)!\over (m!)^N}D(m) z^m
\over\sum_{m\geq0}{(Nm)!\over (m!)^N} z^m},}
our problem then reduces to proving that $f(z^p)-pf(z)\in pz\Z_p[[z]]$ for all
prime $p$. We will first prove that $e^{f(z)}\in1+X\Z_p[[X]]$ for $p\neq N$.
Then show that $f(z^p)-pf(z)\in pz\Z_p[[z]]$ for $p=N$.

\newsec{Step 2: Dwork's Theorems on p-adic hypergeometric series}

We'll not state the most general setting for Dwork's Theorems which require
a rather long discussion. Instead we state the theorems only in the generality
we need here.

{\it Notations:} Let $\Omega$ be the completion of $\bar{\Q_p}$, $\cO$ the ring
of integers of $\Omega$, $\Omega^\times$ the group of units of $\Omega$, $C$
the set of all rational numbers which are p-integral but which are neither zero
nor a negative rational integer. The following are extracted from
\ref\dwork{B.M. Dwork, Ann. Scient. Ec. Norm. Sup. $4^e$ serie, t. 6 (1973)
295-316.}

\theorem{\dwork Let $A_0,A_1:\Z_{\geq0}\ra\Omega^\times$,
$g_0,g_1:\Z_{\geq0}\ra\cO-\{0\}$ be mappings satisfying the conditions that
\item{}(i) $|A_i(0)|=1$;
\item{}(ii) $A_i(m)\in g_i(m)\cO$;
\item{}(iii) for all $a,\mu,s\in\Z_{\geq0}$ such that $a<p$, $\mu<p^s$ we have
\eqn\dumb{{A_0(a+\mu p+mp^{s+1})\over A_0(a+\mu p)}-
{A_1(\mu+mp^s)\over A_1(\mu)}\in p^{s+1}{g_1(m)\over g_0(a+\mu p)}\cO.}
Then for all $m,s,M\in\Z_{\geq0}$, we have
\eqn\dumb{H_a(m,s,M)\in p^{s+1}g_1(m)\cO}
where
\eqn\dumb{H_a(m,s,M)=\sum_{j=mp^s}^{(m+1)p^s-1}
\left(A_0(a+(M-j)p)A_1(j)-A_1(M-j)A_0(a+jp)\right).}
}

Define a mapping $C\ra C$, $x\mapsto x'$, where $x'$ is the element such that
$px'-x$ is the minimal representative in $\Z_{\geq0}$ of the class of
$-x~mod~p$.
Let $\theta_1,..,\theta_N$ be elements of $C$. Define
\eqn\dumb{\eqalign{
F(t)=&\sum_{m\geq0}\alpha(m)t^m\cr
\alpha(m)=&{\Pi_{i=1}^N\theta_i(\theta_i+1)\cdots(\theta_i+m-1)\over
m!^{N-1}}\cr
\tilde{F}(t)=&\sum_{m\geq1}\alpha(m)D(m)t^m\cr
D(m)=&\sum_{i=1}^{N-1} D_{\theta_i}(m)-(N-1)D_1(m)\cr
G(t)=&\sum_{m\geq0}\alpha'(m)t^m\cr
\tilde{G}(t)=&\sum_{m\geq1}\alpha'(m)D'(m)t^m\cr
}}
where $\alpha',D'$ are given by the same formulas as $\alpha,D$ but with the
$\theta_i$ replaced by $\theta_i'$. Note that $F(t)$ is a special case of a
so-called generalized hypergeometric series ${}_{N-1}F_{N-2}[\theta;\sigma;t]$,
ie. it is a power series solution to the generalized hypergeometric equation
\eqn\dumb{\left(t{d\over dt} \Pi_{j=1}^{N-2}(t{d\over dt}
+\sigma_j-1)-t\Pi_{i=1}^{N-1}(t{d\over dt}+\theta_i)\right)g(t)=0.}

\theorem{\dwork In the above notations, we have
\eqn\dumb{{\tilde{G}\over G}(t^p)\equiv p{\tilde{F}\over
F}(t)~~mod~p\Z_p[[t]].}
}

\bs
Now assume that the prime $p\neq N$, and let $\theta_i={i\over N}$.
By $(N,p)=1$, it's easy to check that the mapping $x\mapsto x'$ acts by
permutation on the set $\{\theta_1,..,\theta_{N-1}\}$. This means that
$G=F$ and $\tilde{G}=\tilde{F}$.
We conclude from Theorem 4.1 that for $p\neq N$,
\eqn\dumb{{\tilde{F}\over F}(t^p)\equiv p{\tilde{F}\over
F}(t)~~mod~p\Z_p[[t]].}
By Corollary in section \partI, we have $exp({\tilde{F}(t)\over F(t)})
\in1+t\Z_p[[t]]$. Thus $exp({\tilde{F}(kt)\over F(kt)})
\in1+t\Z_p[[t]]$ for any $k\in\Z_p$.

Now note that $\alpha(m)={(Nm)!\over m!^N N^{Nm}}$, which implies that
$\omega_0(z)= F(zN^N)$, $g(z)=\tilde{F}(zN^N)$.
So for $p\neq N$, we can conclude that $exp({g(z)\over
\omega(z)})\in1+z\Z_p[[z]]$. Thus it remains to prove
$f(z^p)-pf(z)\in pz\Z_p[[z]]$ for $p=N$.

\newsec{Step 3: $p=N$.}

Since ${(Nm)!\over m!^N}$ is a product of binomial coefficients, we have
$\omega_0(z)\in\Z[[z]]$. So it's enough to show that
$\omega_0(z)g(z^p)-p\omega_0(z^p)g(z)\in pz\Z_p[[z]]$. But from eqn. \eqnD,
it's clear that $D_x(m)\in p\Z_p[[z]]$ for $x=\theta_1,..,\theta_{N-1}$. Thus
it's enough to show that (see eqn. \eqng):
\eqn\eqnh{h(z):=\sum A(n)A(m)D_1(m)z^{mp+n}-p\sum A(n)A(m)D_1(n)z^{mp+n}\in
pz\Z_p[[z]]}
where $A(n):={(pm)!\over m!^p}$. The coefficient of $z^{a+Mp}$ in $h(z)$
($0\leq a<p$, $M\in\Z_{\geq0}$) is
\eqn\dumb{L(a+Mp):=\sum_{j=0}^M A(a+jp)A(M-j)\left(D_1(M-j)-pD_1(a+jp)\right).}

{}From eqn. \eqnD, trivially we have $pD_1(a+jp)\equiv D_1(j)~mod~p\Z_p$. Thus
\eqn\dumb{\eqalign{
L(a+Mp)=&\sum_{j=0}^M A(a+jp)A(M-j)\left(D_1(M-j)-D_1(j)\right)\cr
=&\sum_{j=0}^M D_1(j)\left(A(a+jp)A(M-j)-A(a+(M-j)p)A(j)\right)\cr
=&-\sum_{s=0}^r\sum_{m=0}^{p^{1+r-s}-1}Y_{m,s}~~~~{\rm where}\cr
Y_{m,s}=&\left(D_1(mp^s)-D_1([{m\over p}]p^{s+1})\right) H_a(m,s,M)\cr
H_a(m,s,M)=&\sum_{j=mp^s}^{(m+1)p^s-1}
\left(A(a+(M-j)p)A(j)-A(M-j)A(a+jp)\right).
}}
for some $r$ with $p^r>M$. The last expression for $L(a+Mp)$ is obtained by
applying Lemma 4.2 in \dwork. Now by writing $m=b+Rp$, $0\leq b<p$,
$R=[{m\over p}]$ the integer part of $m\over p$, we have
\eqn\dumb{
D_1(mp^s)-D_1([{m\over p}]p^{s+1})=
(1+{1\over 2}+\cdots +{1\over Rp^{s+1}+bp^s})-
(1+{1\over 2}+\cdots +{1\over Rp^{s+1}})\equiv0~~mod~{1\over p^s}\Z_p.}
Hence
\eqn\dumb{Y_{m,s}\in {1\over p^s}H_a(m,s,M)\Z_p.}
To complete the proof that $L(a+Mp)\in p\Z_p$, we'll prove that
\eqn\dumb{H_a(m,s,M)\in p^{s+1}\Z_p}
for $a,m,s,M\in\Z_{\geq0}$, $0\leq a<p$.
For this we'll apply Theorem 1.1 with $A_0(n)=A_1(n)=A(n)$, $g_0(n)=g_1(n)=1$.
The hypotheses (i), (ii) there obviously hold. So we must check hypothesis
(iii). Thus we claim that
for all $a,\mu,s\in\Z_{\geq0}$ such that $a<p$, $\mu<p^s$ we have
\eqn\dumb{ord\left({A(a+\mu p+mp^{s+1})\over A(a+\mu p)}-
{A(\mu+mp^s)\over A(\mu)}\right)\geq s+1.}
To prove this we will use the p-adic Gamma function to estimate the LHS.

First we state some simple facts. Recall that
\eqn\dumb{ord~A(n)=ord~(pn)!-p~ord~n!=S(n)}
where $S(n)$ is the sum of the p-adic digits of $n$. Recall that the p-adic
Gamma function is given by
\eqn\dumb{\eqalign{
\Gamma_p(n)=&(-1)^n\gamma(n)~~~~{\rm where}\cr
\gamma(n)=&\Pi_{0<j<n,~(n,p)=1} j.
}}
We can write $\gamma(1+np)={(np)!\over p\cdot 2p\cdots np}$, hence
\eqn\npgamma{(np)!=\gamma(1+np)n!p^n.}
Also for positive integers $n,k,r$, we have (see \ref\lang{S. Lang, {\sl
Cyclotomic Fields I and II}, Springer Verlag. See p314.})
\eqn\dumb{\Gamma_p(n+kp^r)\equiv\Gamma_p(n)~~mod~p^r.}
We apply these formulas repeatedly in the following computation:
\vfill\eject

\eqn\dumb{\eqalign{
&{A(a+\mu p+mp^{s+1})A(\mu)\over A(a+\mu p)A(\mu+mp^s)}\cr
=&{(ap+\mu p^2+mp^{s+2})!\over(a+\mu p+mp^{s+1})!^{p}}
{(\mu+mp^s)!^p\over(\mu p+mp^{s+1})!}
{(a+\mu p)!^p\over(ap+\mu p^2)!}
{(\mu p)!\over \mu!^p}\cr
=&\gamma(1+ap+\mu p^2+mp^{s+2})(a+\mu p+mp^{s+1})!^{1-p}p^{a+\mu p+mp^{s+1}}\cr
&\times\gamma(1+\mu p+mp^{s+1})^{-1}(\mu +mp^{s})!^{p-1}p^{(\mu+mp^{s})(-1)}\cr
&\times\gamma(1+ap+\mu p^2)^{-1}(a +\mu p)!^{p-1}p^{(a+\mu p)(-1)}\cr
&\times\gamma(1+\mu p)\mu!^{1-p}p^\mu\cr
=&\gamma(1+ap+\mu p^2+mp^{s+2})\Pi_{i=1}^a(i+\mu p+mp^{s+1})^{1-p}\cr
&\times\gamma(1+\mu p+mp^{s+1})^{1-p} (\mu+mp^s)!^{1-p} p^{(\mu+mp^s)(1-p)}
p^{a+\mu p+mp^{s+1}}\cr
&\times\gamma(1+\mu p+mp^{s+1})^{-1}(\mu +mp^{s})!^{p-1}p^{(\mu+mp^{s})(-1)}\cr
&\times\gamma(1+ap+\mu p^2)^{-1}\Pi_{i=1}^a(i +\mu p)^{p-1}
\gamma(1 +\mu p)^{p-1} \mu!^{p-1} p^{\mu(p-1)} p^{(a+\mu p)(-1)}\cr
&\times\gamma(1+\mu p)\mu!^{1-p}p^\mu\cr
=&{\gamma(1+ap+\mu p^2+mp^{s+2})\gamma(1 +\mu p)^p\over
\gamma(1+\mu p+mp^{s+1})^p\gamma(1+ap+\mu p^2)}
\Pi_{i=1}^a(i+\mu p+mp^{s+1})^{1-p}\Pi_{i=1}^a(i +\mu p)^{p-1}\cr
\equiv&{\gamma(1+ap+\mu p^2+mp^{s+2})\gamma(1 +\mu p)^p\over
\gamma(1+\mu p+mp^{s+1})^p\gamma(1+ap+\mu p^2)}
(\Pi_{i=1}^a(i+\mu p)+O(p^{s+1}))^{1-p}\Pi_{i=1}^a(i +\mu p)^{p-1}\cr
\equiv&{\gamma(1+ap+\mu p^2+mp^{s+2})\gamma(1 +\mu p)^p\over
\gamma(1+\mu p+mp^{s+1})^p\gamma(1+ap+\mu p^2)} (1+O(p^{s+1}))\cr
=&{\Gamma_p(1+ap+\mu p^2+mp^{s+2})\Gamma_p(1 +\mu p)^p\over
\Gamma_p(1+\mu p+mp^{s+1})^p\Gamma_p(1+ap+\mu p^2)} (1+O(p^{s+1}))\cr
=&{(\Gamma_p(1+ap+\mu p^2)+O(p^{s+2}))\Gamma_p(1 +\mu p)^p\over
\Gamma_p(1+ap+\mu p^2)(\Gamma_p(1+\mu p)+O(p^{s+1}))^p} (1+O(p^{s+1}))\cr
\equiv&1+O(p^{s+1})
}}
Thus we conclude that
\eqn\dumb{\eqalign{
&ord\left({A(a+\mu p+mp^{s+1})\over A(a+\mu p)}-
{A(\mu+mp^s)\over A(\mu)}\right)\cr
=&ord~{A(\mu+mp^s)\over A(\mu)}+ord\left({A(a+\mu p+mp^{s+1})A(\mu)\over
A(a+\mu p)A(\mu+mp^s)}-1\right)\cr
\geq&S(\mu)+S(m)-S(\mu)+s+1\geq s+1.
}}
This completes our proof.

\newsec{Other cases and applications}

It turns out that the above technique for studying the integrality property
for the degree $p$ hypersurfaces in $\P^{p-1}$ is also applicable in
other cases with some minor modifications. For example, it can be verified
that in numerous cases of Calabi-Yau threefolds, such a technique applies.

To see how this is done, let's consider for example the case of the degree
10 hypersurface in $\P^4[1,1,1,2,5]$. The Picard-Fuchs equation in this case
is given by
\eqn\dumb{
\left(\Theta_z^4  - 80z (1 + 10 \Theta_z) (3 + 10 \Theta_z)
(7 + 10 \Theta_z) (9 + 10 \Theta_z)\right)f=0.}
First we rescale $z$ by $z\mapsto z/80\cdot 10^4$. The equation becomes
\eqn\dumb{
\left(\Theta_z^4  - z ({1\over 10} + \Theta_z) ({3\over 10}+ \Theta_z)
({7\over 10}+\Theta_z) ({9\over 10}+ \Theta_z)\right)f=0.}
Then the arguments in Parts I and II above apply for all $p\neq 2,5$ and with
the data $\theta_1={1\over 10}, \theta_2={3\over 10},
\theta_3={7\over 10},\theta_4={9\over 10}$.

To treat $p=2,5$, we apply a similar argument as in Step 3 above together
with an estimate on the coefficients of an appropriate generalized
hypergeometric series using the $p-adic$ gamma function. This argument
goes through without a hitch, and hence prove that the mirror map in this
case is integral.

Finally we note that the integrality property of the mirror map is
crucial for us in our previous study on some arithmetic properties
of the quantum Yukawa coupling \ly. We plan to further investigate these
properties.

\listrefs
\end